\documentclass[journal]{IEEEtran}
\usepackage{amsmath,amsfonts}
\usepackage{algorithmic}
\usepackage{algorithm}
\usepackage{array}
\usepackage[caption=false,font=normalsize,labelfont=sf,textfont=sf]{subfig}
\usepackage{textcomp}
\usepackage{stfloats}
\usepackage{url}
\usepackage{verbatim}
\usepackage{graphicx}
\usepackage{cite}
\usepackage{orcidlink}
\usepackage{flushend}
\begin{document}

\title{An Integrated Deep-Cryogenic Temperature Sensor in CMOS Technology for Quantum Computing Applications}

\author{Fabio Olivieri\,\orcidlink{0009-0000-0278-8100}, Grayson M. Noah\,\orcidlink{0009-0000-1110-1251}, Thomas Swift\,\orcidlink{0009-0005-1318-6865}, M. Fernando Gonzalez-Zalba\,\orcidlink{0000-0001-6964-3758}, \\ 
John J. L. Morton\,\orcidlink{0000-0002-0891-1111}, and Alberto Gomez-Saiz\,\orcidlink{0000-0001-5902-0207}
        % <-this % stops a space
\thanks{Fabio Olivieri (\href{mailto:fabio@quantummotion.tech}{fabio@quantummotion.tech}), Grayson M. Noah (\href{mailto:grayson@quantummotion.tech}{grayson@ quantummotion.tech}), M. Fernando Gonzalez-Zalba, and Alberto Gomez-Saiz are with Quantum Motion, 9 Sterling Way, London, N7 9HJ, United Kingdom.} % <-this % stops a space
\thanks{Thomas Swift and John J. L. Morton are with Quantum Motion, 9 Sterling Way, London, N7 9HJ, United Kingdom, and also with the London Centre for Nanotechnology, UCL, London, WC1H 0AH, United Kingdom.} \\
\thanks{Manuscript received August 30, 2024; revised December 20, 2024.}}

% The paper headers
\markboth{ } %IEEE draft %Journal of TBU Class Files,~Vol.~14, No.~8, August~2021} %TBU
{Shell \MakeLowercase{\textit{et al.}}: A Sample Article Using IEEEtran.cls for IEEE Journals}

\IEEEpubid{} %0000--0000/00\$00.00~\copyright~2021 IEEE} %TBU

% Remember, if you use this you must call \IEEEpubidadjcol in the second column for its text to clear the IEEEpubid mark.

\maketitle

\begin{abstract}
On-chip thermometry at deep-cryogenic temperatures is vital in quantum computing applications to accurately quantify the effect of increased temperature on qubit performance. In this work, we present a sub-1~K temperature sensor in CMOS technology based on the temperature dependence of the critical current of a superconducting (SC) thin-film. The sensor is implemented in 22-nm fully depleted silicon on insulator (FDSOI) technology and comprises a 6-nA-resolution current-output digital-to-analog converter (DAC), a transimpedance amplifier (TIA) with a SC thin-film as a gain element, and a voltage comparator. The circuit dissipates 1.5~\textmu W and is demonstrated operating at ambient temperatures as low as 15~mK, providing a variable temperature resolution reaching sub-10~mK. 
\end{abstract}


\section{Introduction}
\IEEEPARstart{E}{lectron} spin qubit fidelities in silicon begin to degrade as their local temperature rises, with demonstrated readout fidelity dropping below 99\% above 1~K \cite{huang2023high}. Below 1~K, the limited cooling power at the mixing chamber (MXC) plate of dilution refrigerators \cite{pobell2007matter} and the reduced thermal conductivity and heat capacity of silicon and oxides compared to room temperature \cite{slack1964thermal} make low-power circuitry and local on-chip thermometry essential to characterizing and operating qubits on a CMOS chip. Thermometry based on gate resistance sensing has been used for characterization of transistor self-heating and cross-device heating but has been limited to local temperatures well above 1~K \cite{10019322,10449894}. For lower local temperatures which are relevant to the operation of qubits, primary thermometry methods like Coulomb blockade and quantum dot thermometry have been used \cite{Kauppinen1998, dekruijf2023qdt,maradan}, but such techniques typically require complex data post-processing involving fitting functions, making them less suitable for on-chip integration. On-chip diode thermometry has been demonstrated at sub-1~K temperatures \cite{noah2023cmos}. However, sub-1~K diode-based thermometry requires high-resolution measurement circuitry due to the small temperature sensitivity of the diode's electrical characteristics in this regime.

Materials used in the fabrication of poly-silicon resistors and the transistor gate stack in some CMOS processes exhibit superconductive behaviour near and below 1~K \cite{noah2023cmos}. Structures containing superconductive material exhibit a large change in electrical characteristics when the material undergoes a temperature-induced phase transition. This property can be exploited to generate a large signal change, reducing the resolution requirements for the measurement circuitry and hence enabling the implementation of low-complexity and low-power sub-1~K temperature sensing circuits.

Here, we present a local temperature-sensing system fabricated in GlobalFoundries 22FDX that exploits the temperature dependence of the critical current that induces the normal-to-SC transition (retrapping current $I_{\textnormal{RT}}$) of a SC-capable poly-silicon resistor structure. The presented work integrates a SC element, a bias generator circuit, and a comparator circuit and is suitable for monolithic integration with CMOS-compatible qubits.

\section{Circuit Design and Operation}
\IEEEpubidadjcol

\begin{figure*}[t]
\centering
\includegraphics[width=0.95\linewidth]{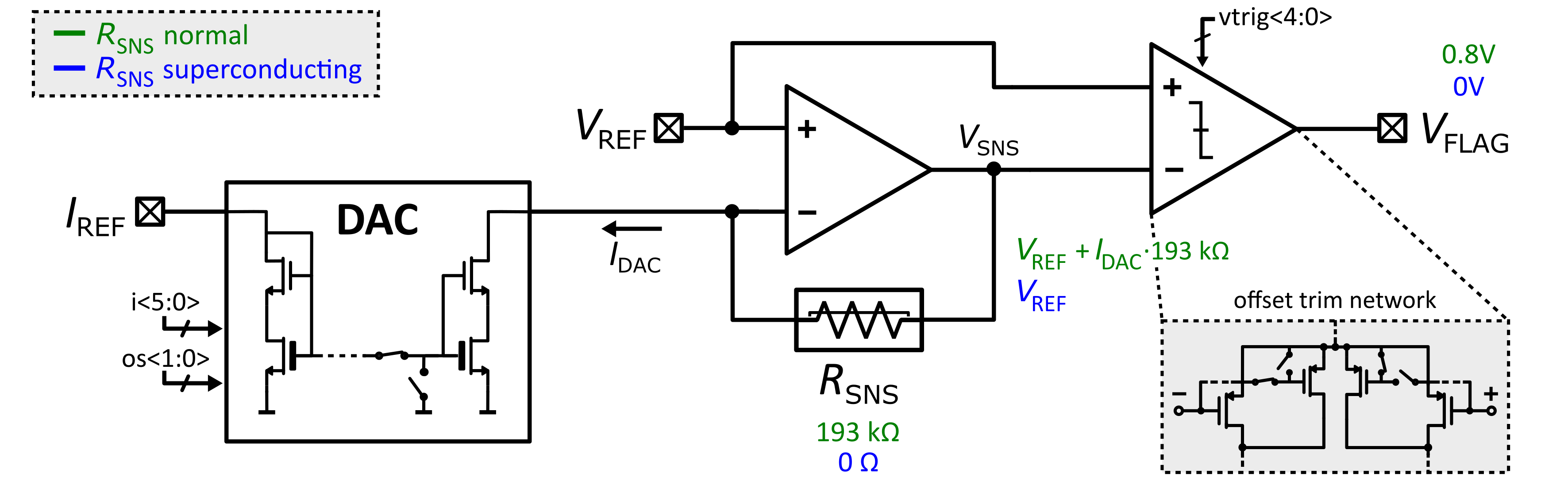}
\caption{Simplified schematic of the SC-based temperature sensor. All circuits are powered by a 0.8-V supply.}
\label{fig1}
\end{figure*}

The simplified schematic of the temperature sensor is shown in Fig. \ref{fig1}, and the circuit operates from a 0.8-V supply. The current-output digital-to-analog converter (DAC) generates a bias current $I_{\textnormal{DAC}}$ which is applied to the SC-capable resistor $R_{\textnormal{SNS}}$ to generate the SC-state-dependent sensing voltage $V_{\textnormal{SNS}}$. This voltage is then compared to the reference voltage $V_{\textnormal{REF}}$ by the voltage comparator to produce the logic output $V_{\textnormal{FLAG}}$.
By sweeping the DAC code, it is possible to induce a state transition in the SC-capable resistor and determine the value of $I_{\textnormal{RT}}$ by monitoring the comparator output voltage $V_{\textnormal{FLAG}}$ value, as depicted in Fig. \ref{fig1}. 

The DAC circuit is implemented using a 6-bit programmable-gain current mirror, which employs self-cascode unit cells to provide high output impedance while operating within a limited voltage headroom \cite{yan2000low}. 
Internally, the DAC adopts a segmented architecture \cite{lin199810}, where the three most significant bits (MSBs) use thermometer decoding, while the three least significant bits (LSBs) are binary-weighted. 
The DAC resolution is set by an external reference current $I_{\textnormal{REF}}$ (nominally 50~nA), which gives an LSB equal to $I_{\textnormal{REF}}/8$.
The DAC output range can be shifted using a 2-bit programmable offset with a unit step size of 63 LSBs. This feature effectively extends the full-scale range up to $4 \cdot 63 \cdot I_{\textnormal{REF}}/8$, divided into 4 sections of 6-bit resolution. 
When transitioning between sections via the offset control, the DAC is most sensitive to mismatch effects which cause linearity degradation and non-monotonicity. To address this issue, adjacent sections are designed to overlap by 1~LSB, ensuring that the transition between the top of one section and the bottom of the next can be made with an output error of less than 1~LSB.

The TIA circuit employs a two-stage Miller-compensated OTA and uses an external reference voltage $V_{\textnormal{REF}}$ = 0.4~V for biasing the positive input terminal and setting the input common-mode voltage. This same voltage $V_{\textnormal{REF}}$ also defines the DAC output voltage and is used to bias the positive input of the comparator.
The SC-capable resistor closes the negative feedback loop around the OTA, generating the sensing voltage $V_{\textnormal{SNS}}$ on the comparator negative input.

The voltage comparator circuit uses a high-gain operational amplifier (op-amp) in open-loop configuration followed by a logic inverter to generate the logic output $V_{\textnormal{FLAG}}$. The op-amp includes a 5-bit offset trim network to calibrate the input offset voltage across process and temperature variations. As depicted in Fig. \ref{fig1}, offset calibration is achieved by introducing a size mismatch between the transistors forming the differential input pair of the op-amp.

The circuit elements are digitally controlled through an on-chip register which is interfaced externally via a JTAG test access port (TAP). An additional test mode is included to enable Kelvin connections to each of the two terminals of $R_{\textnormal{SNS}}$, allowing isolated testing of the different circuit blocks.

\begin{figure}[H]
\centering
\includegraphics[width=0.965\linewidth]{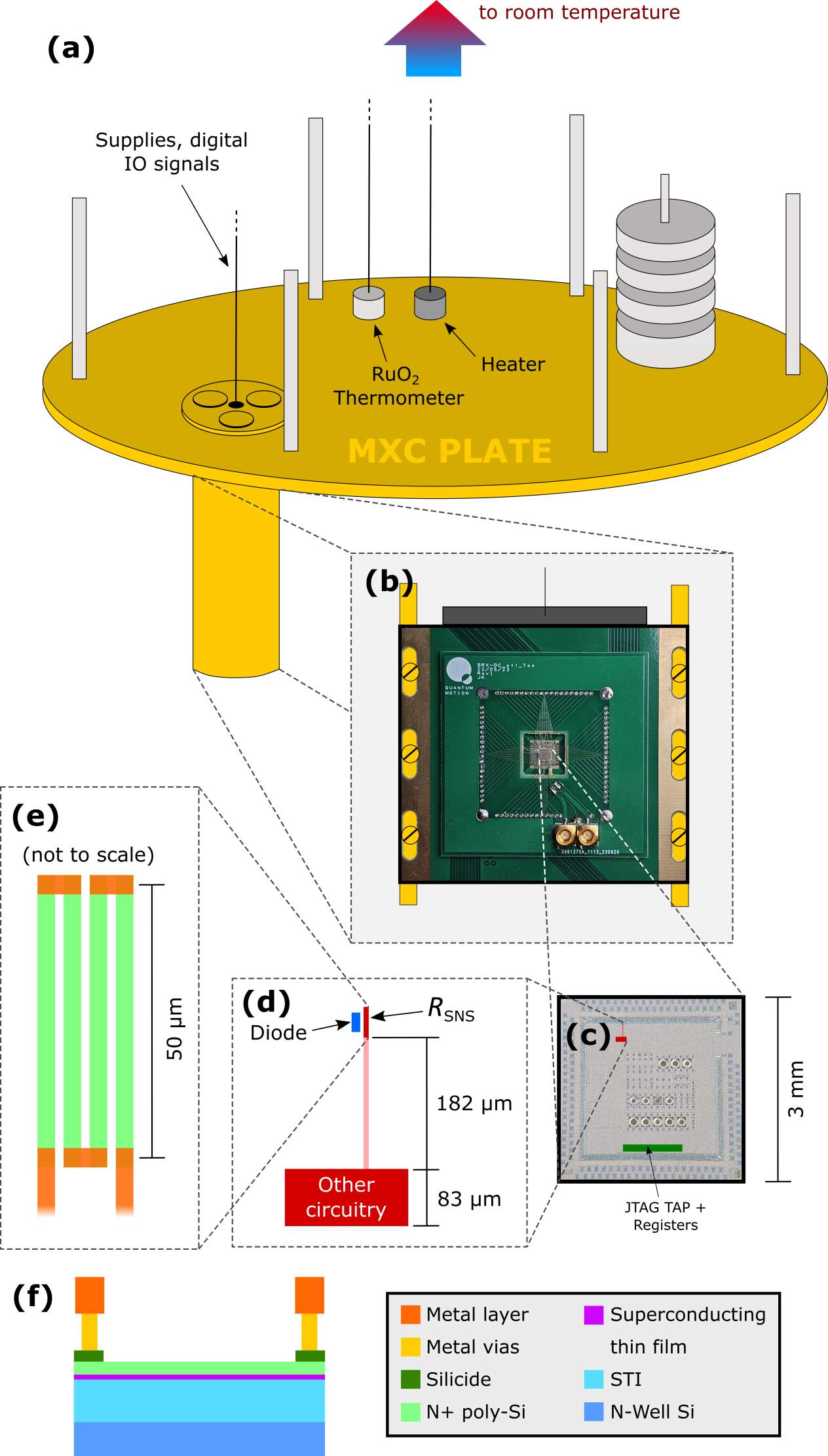}
\caption{Measurement setup and sensor implementation details. (a) Diagram of mixing chamber plate of the dilution refrigerator used for measurements. (b) Rail-mounted PCB used for measurements. (c) Micrograph of chip top surface showing circuitry positions. (d) Sensor layout. (e) Top-view diagram (not to scale) showing the topology used to connect the four resistor units in series to form $R_{\textnormal{SNS}}$. (f) Cross-section diagram (not to scale) of one of the resistor units detailing the constituent materials.}
\label{fig2}
\end{figure}

As shown in Fig. \ref{fig2}e, the SC-capable resistor $R_{\textnormal{SNS}}$ is composed of four identical units connected in series, resulting in a total resistance of 193~k$\Omega$ in the normal state at deep cryogenic temperatures. The number of identical units connected in series minimally affects the critical current values, which are primarily dependent on the width of the units.
In the SC state, the resistance of the thin-film layer becomes zero, and $R_{\textnormal{SNS}}$ reduces to a few tens of ohms due only to residual routing and contact resistance.

The voltage drop across $R_{\textnormal{SNS}}$ in the SC state with the maximum DAC code is always smaller than the voltage drop in the normal state with the minimum DAC code.
This allows the comparator offset trim to be calibrated such that $V_{\textnormal{FLAG}}$ is 0~V (logic 0) when $R_{\textnormal{SNS}}$ is in the SC state and 0.8~V (logic 1) when $R_{\textnormal{SNS}}$ is in the normal state, regardless of $I_{\textnormal{DAC}}$. %, if the introduced offset is small enough not to limit the sensor sensitivity at low DAC codes.
After calibration, the following procedure is used to measure the value of the retrapping current $I_{\textnormal{RT}}$ of $R_{\textnormal{SNS}}$. First, $I_{\textnormal{DAC}}$ is set to a current level known to bias $R_{\textnormal{SNS}}$ in the normal state and produce an observed stable logic 1 on $V_{\textnormal{FLAG}}$. Then, the DAC code is stepped down one code at the time until $V_{\textnormal{FLAG}}$ transitions to logic 0, at which point $I_{\textnormal{RT}}$ = $I_{\textnormal{DAC}}$.

\section{Results}

\begin{figure}[tb]
\centering
\includegraphics[width=0.95\linewidth]{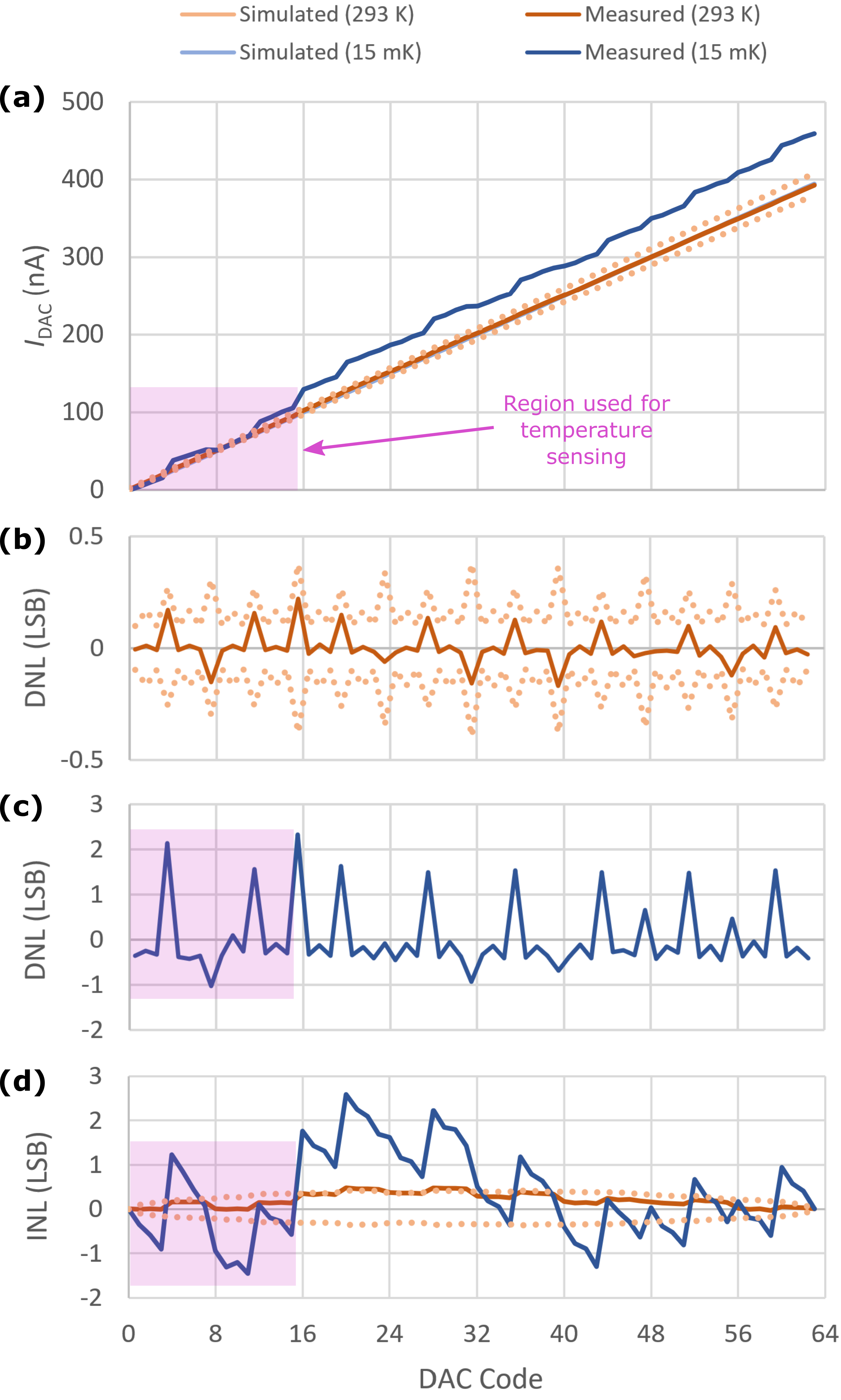}
\caption{DAC simulated and measured results at 293~K and 15~mK ambient temperature. Dotted lines represent $\pm$3$\sigma$ bounds from Monte Carlo simulation at 293~K. (a) DAC transfer characteristic; (b-d) DAC linearity.}
\label{fig3}
\end{figure}

\begin{figure}[tb]
\centering
\includegraphics[width=0.95\linewidth]{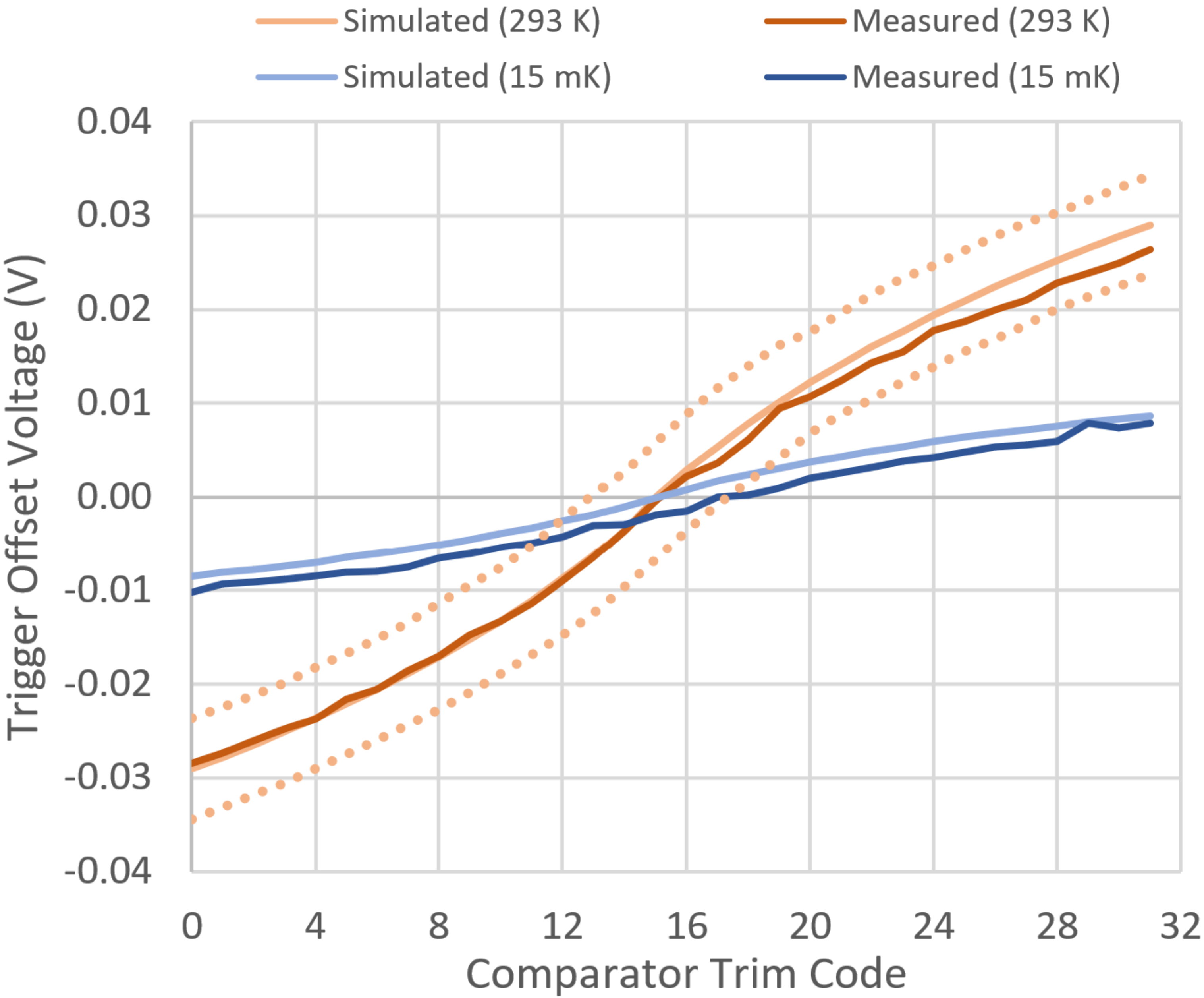}
\caption{Simulated and measured comparator trigger voltage across offset trim codes at 293~K and 15~mK ambient temperature. Dotted lines represent $\pm$3$\sigma$ bounds from Monte Carlo simulation at 293~K.}
\label{fig4}
\end{figure}

The fabricated die is wire-bonded to a PCB which is mounted to the MXC stage of a dilution refrigerator as shown in Fig. \ref{fig2}. We show the simulated and measured results at room temperature (293 K ambient) and deep-cryogenic temperature (15 mK ambient) for the DAC in Fig. \ref{fig3} and the comparator in \ref{fig4}. The cryogenic simulations use custom BSIM-IMG v102.9.6 \cite{bsimimg} model cards extracted using cryogenic measurement data from large arrays of FETs of various sizes \cite{eastoe2024efficient}. Cryogenic variability is not modeled here, so only typical cryogenic simulations are included. Typical simulations are not very meaningful for linearity of the DAC (and are not plotted on Fig. \ref{fig3}b-d) since mismatch is the primary source of non-linearity in this design. The pattern of non-linearity peaks every four codes is typical of this kind of DAC design due to its 3 + 3 segmented architecture with the 3 LSBs binary weighted. While the measured DAC linearity is near simulated $\pm$3$\sigma$ bounds at room temperature, it suffers major degradation at cryogenic temperatures which is suspected to be due to worse FET mismatch \cite{hart2020}. The addition of local mismatch derived from bulk cryogenic characterization data to the custom model cards is essential to verify this assumption, as the differential non-linearity (DNL) peaks are nearly an order of magnitude higher at cryogenic temperature (Fig. \ref{fig1}b and \ref{fig1}c are thus plotted separately for clarity). 

The comparator trigger voltage trim behaves closely to simulations at room temperature. The simulation at cryogenic temperature also demonstrates an accurate prediction of the behavior other than a small offset assumed to be due to process variation, which is not included in this simulation. 

\begin{figure}[!t]
\centering
\includegraphics[width=\linewidth]{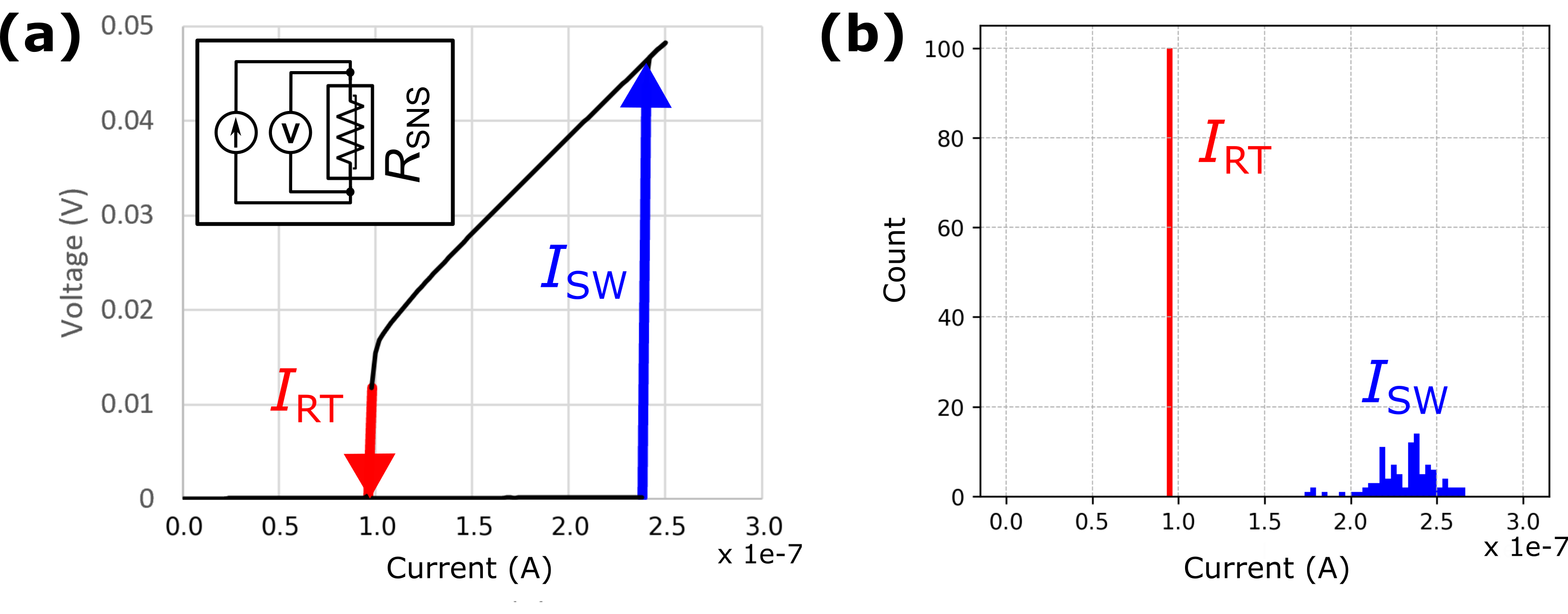}
\caption{Externally measured voltage-current characteristics of the SC-capable resistor $R_{\textnormal{SNS}}$. (a) Hysteretic I-V curve of $R_{\textnormal{SNS}}$ near 400~mK with retrapping and switching currents noted (inset: simplified measurement circuit diagram). (b) Histograms of $R_{\textnormal{SNS}}$ state transition currents.}
\label{fig5}
\end{figure}

Although the circuit is also capable of detecting the SC-to-normal (switching) transition of $R_{\textnormal{SNS}}$ which would exhibit greater sensitivity (in A/K), the switching current $I_{\textnormal{SW}}$ that induces this transition is subject to inherent stochastic variation \cite{bardeen1962critical} and would require averaging over multiple measurements for a reliable estimation (Fig. \ref{fig5}b). Thus, the more stable retrapping current $I_{\textnormal{RT}}$ is used.

\begin{figure}[!t]
\centering
\includegraphics[width=0.95\linewidth]{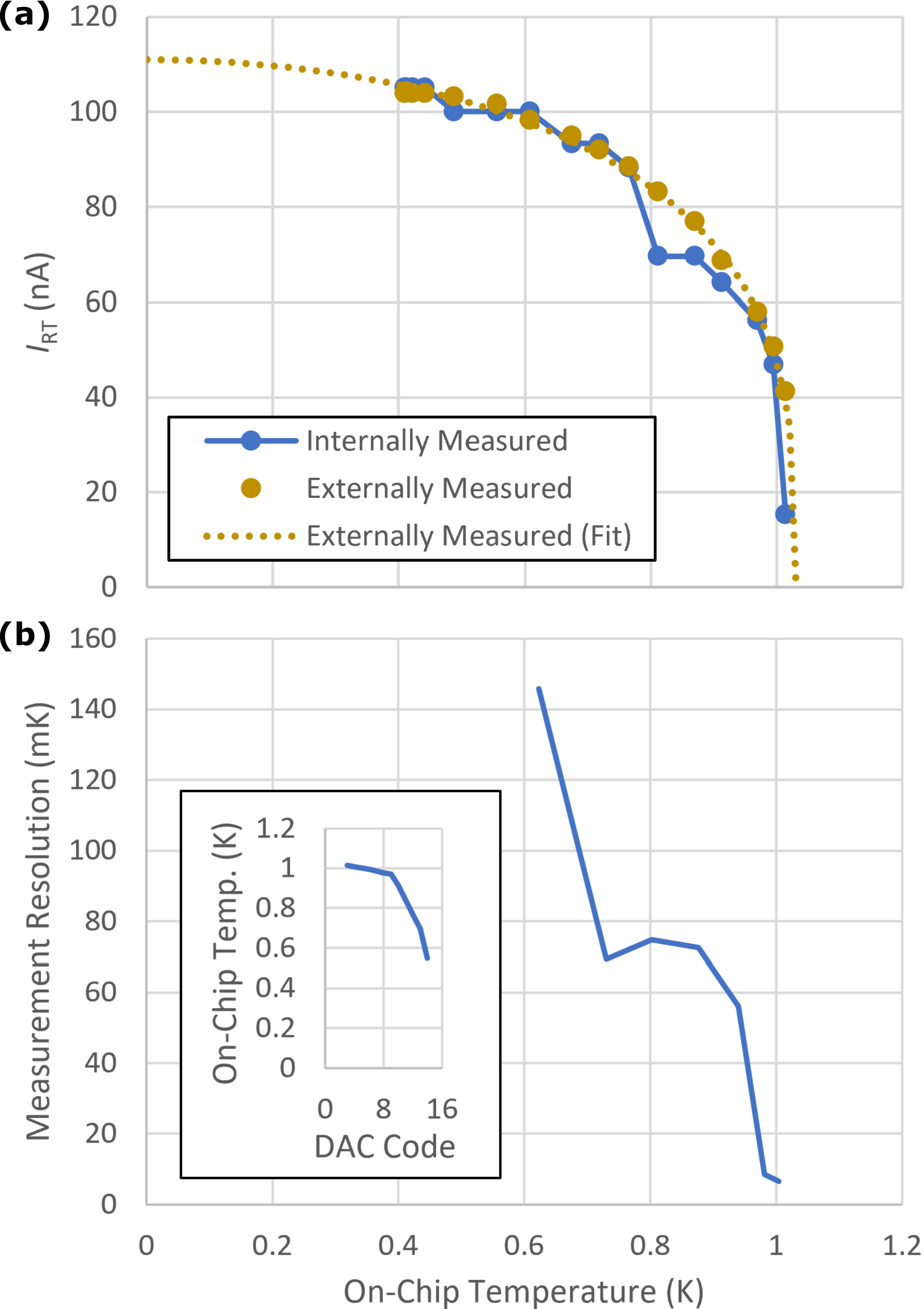}
\caption{Thermometry results. (a) External and integrated measurements of $I_{\textnormal{RT}}$ of $R_{\textnormal{SNS}}$ across temperature. (b) Approximate resolution of integrated temperature measurement with inset code-temperature map.}
\label{fig6}
\end{figure}

To validate the performance of the SC-based temperature sensor, an on-chip diode with dedicated measurement pads has been positioned in close proximity to $R_{\textnormal{SNS}}$ as shown in Fig. \ref{fig2}d. The diode is first calibrated over a range of MXC temperatures with the rest of circuitry powered off as described in \cite{noah2023cmos}. During this calibration, the only power dissipated on the chip is due to the Joule heating of the diode sensor (apx. 1 nW) which is low enough to maintain sensitivity throughout the temperature range of interest. Then when circuitry is powered up and $I_{\textnormal{RT}}$ is measured at different stable MXC temperatures (as controlled by a separate heater and ruthenium oxide thermometer located on the MXC plate), the calibrated diode is measured to provide the local temperature reference (x-axis) in Fig. \ref{fig6}. In addition to the 1.5~\textmu W dissipated by the circuitry shown in Fig. \ref{fig1}, there is approximately 3~\textmu W of static power dissipated in other areas of the chip related to digital and other auxiliary circuitry, leading to the minimum detected on-die temperature being $\sim$400~mK.

External measurements (using the test mode) of $I_{\textnormal{RT}}$ are performed with all circuit elements disabled and are thus less susceptible to transient heating effects from on-chip sources. The critical temperature of $R_{\textnormal{SNS}}$ is observed to be apx. 1.03 K, above which the sensor is not useful. $I_{\textnormal{RT}}$ measured by the integrated circuit matches external measurement very closely other than a deviation near 0.8~K corresponding to the high DNL in the transition between codes 11 and 12. This high DNL results in a sudden drop in the accuracy of the measured $I_{\textnormal{RT}}$ since intermediate currents cannot be sourced to determine $I_{\textnormal{RT}}$ more precisely. Additionally, during this code transition the larger current step change causes a larger overshoot and undershoot, with the latter making more likely a false trigger of the retrapping event. Similarly high DNL occurs in the transition between codes 3 and 4. However, this code transition is only used at higher temperatures at which the temperature sensitivity of $I_{\textnormal{RT}}$ (in A/K) is very low and hence its effect on the degradation of $I_{\textnormal{RT}}$ measurement accuracy is less apparent.
 
To determine the temperature resolution of the sensor, we record at several temperatures the DAC code that triggers the retrapping event and compare to the local temperature as measured by the on-chip diode (see inset of Fig. \ref{fig6}b). For codes that do not trigger a retrapping event in the temperatures of our test set, we interpolate between codes for which data exists. The magnitude of the slope of this mapping represents the implied temperature difference between two adjacent codes (the minimum resolvable temperature difference) as plotted across temperature in the main plot of Fig. \ref{fig6}b. The resulting temperature resolution varies from $\sim$150~mK at 0.6~K to $\sim$7~mK at 1~K.

\section{Conclusion}
The sensor is effective within the 0.6-1~K range, which aligns with the temperature range at which spin qubit fidelity begins to degrade. In addition to its use for sub-1~K thermometry of cryo-CMOS systems, the presented thermometry solution could be deployed in future quantum systems that monolithically integrate electronics and qubits. Proposals for such systems offer scalability benefits but risk thermally induced degradation of qubits' quantum states due to cross-heating from the integrated electronics \cite{mfgz_natelec}. In a hypothetical monolithically integrated system located on the MXC plate of a dilution refrigerator, the electronics can dissipate at least a few tens of \textmu W RMS power while keeping the qubit temperatures on the same chip below 1 K \cite{noah2023cmos}. The presented temperature sensing circuit dissipates $\sim$1.5 \textmu W and hence could be accommodated within the power budget of such system, providing a way to monitor any localized cross-heating that may occur.
The total area of the circuit is $<$ 0.015~mm\textsuperscript{2}, which along with its low power dissipation enables placement of multiple copies of the temperature sensor in the same chip to monitor different temperature-sensitive sections.
Additionally, a single DAC can be shared via time multiplexing to multiple sensing elements, further reducing area and static power dissipation per sensor. 

One limitation of the presented approach is that the $I_{\textnormal{DAC}}$ current sweep used to determine $I_{\textnormal{RT}}$ is slower than performing a single point measurement. For the most time-critical thermal sensing applications, a different technique can be used with the same circuitry to implement a fast single-temperature threshold detector; in this mode of operation, a constant current bias can be selected to be applied to $R_{\textnormal{SNS}}$ while in the SC state such that the switching transition will be triggered if the temperature rises above a certain threshold at which a specific protective action is desired to occur.

The presented demonstration of the use of the transition current of a SC element in a circuit implemented in a standard CMOS process suggests that other properties of superconductors (e.g. kinetic inductance) could also be exploited in future cryo-CMOS circuits.

\section*{Acknowledgments}
The authors acknowledge Charan Kocherlakota, Debargha Dutta, James Kirkman, and Jonathan Warren of Quantum Motion for their technical support. T. S. acknowledges the Engineering and Physical Sciences Research Council (EPSRC) through the Centre for Doctoral Training in Delivering Quantum Technologies [EP/S021582/1].  M. F. G.-Z. acknowledges a UKRI Future Leaders Fellowship [MR/V023284/1].

\bibliographystyle{IEEEtran}
\bibliography{references}

\end{document}